\documentclass[11pt,a4paper]{article}
\usepackage{jheppub}

\subheader{\today}
\title{Testing holographic computation of entanglement pseudo-entropy in dS\textsubscript{3}/ICFT\textsubscript{2}}
\author[a]{Liang Li}
\affiliation[a]{Center for Gravitational Physics and Quantum Information,\\
Yukawa Institute for Theoretical Physics, Kyoto University,\\
Kitashirakawa Oiwakecho, Sakyo-ku, Kyoto 606-8502, Japan}
\emailAdd{li.liang@yukawa.kyoto-u.ac.jp}
\abstract{
We establish a bottom-up model for studying the `would-be' de-Sitter/interface CFT (dS/ICFT) correspondence on the gravity side, using the Coleman-De Luccia instanton. After absorbing the region where the inflaton changes dramatically by a brane, the spacetime consists of two de Sitter spacetimes with different radii joined by a brane, which is expected to be dual to an interface CFT provided that the dS/CFT conjecture is correct. Our goal is to analyze existing prescriptions for the holographic computation of pseudo-entropy, i.e. the extremal surface in such a spacetime. We argue that some of them would cause parodox with the results expected from the universal properties of interface CFT and compare our results with that of Anti de Sitter/interface CFT correspondence (AdS/ICFT) after analytic continuation. We also read off the $g$-function from the entanglement entropy associated with a subregion perpendicular to the interface.
}

\usepackage{indentfirst}
\usepackage{amsmath}
\usepackage{amssymb}
\usepackage{amsfonts}
\usepackage{amsthm}
\usepackage{amscd}
\usepackage{ascmac}
\usepackage{bm}

\numberwithin{equation}{section}

\usepackage{url}
\usepackage{multirow}
\usepackage{amsfonts,amssymb}
\usepackage{float}
\usepackage{graphicx}
\usepackage{mathtools}
\usepackage{slashed}
\usepackage{empheq}
\usepackage{tikz}
\usepackage{enumitem}

\numberwithin{equation}{section}

\begin{document}

\begin{flushright}
YITP-26-114\\
\end{flushright}

\maketitle
\clearpage

\section{Introduction}
The correspondence between anti de Sitter spacetime and conformal field theory, the so-called AdS/CFT correspondence \cite{Maldacena:1997,GKP:1998,Witten:1998}, provides a well-controlled framework to study quantum gravity in anti de-Sitter spacetime with a negative cosmological constant. Later, a holographic prescription for computing the entanglement entropy in conformal field theories \cite{RT:2006} and its time-dependent generalization \cite{HRT:2007} was proposed. According to this prescription, the entanglement entropy of a subregion in CFT is proportional to the area of the codimension 2 extremal surface in the bulk which is homologous to the subregion. For a compact and connected subregion, the extremal surface has a turning point in the bulk, which plays a crucial role in entanglement wedge reconstruction \cite{DongHarlowWall:2016,AlmheiriDongHarlow:2014}.

Despite the success of AdS/CFT correspondence, the proposed dS/CFT correspondence \cite{Strominger:2001,Maldacena:2002} which is of more phenomenological interest remains mysterious. The dual theory includes primaries with complex conformal dimension and has imaginary central charge, hence is supposed to be non-unitary. Moreover, the entanglement entropy in the dual CFT takes complex values and has been interpreted as pseudo-entropy \cite{Nakata:2020}. The pseudo-entropy is supposed to be computed by an extremal surface in the bulk de Sitter spacetime which is timelike and has no turning point in the Lorentzian dS spacetime \cite{Narayan:2015,Narayan:2017}, hence leading to a puzzle for the entanglement wedge reconstruction. There are several proposals trying to give a prescription for the extremal surface in dS:

\begin{enumerate}[label=\textbf{Prescription \arabic*.},leftmargin=*]
    \item Prepare the Hartle-Hawling no boundary state and connect the timelike extremal surface in the Lorentzian spacetime `smoothly' to the spacelike part in the Euclidean spacetime \cite{Doi:2022PRL,Narayan:2022};
    \item Extend the integral contour which computes the area of extremal surface to complex plane while keeping the timelike part on the real axis\cite{Heller:2024whi,Heller:2025kvp,FujikiKoharaShinmyoSuzuki:2025};
    \item Compute the area of the extremal surface using complexified coordinates \cite{Narayan:2015}.
\end{enumerate}
We explain the motivations behind these prescriptions here. The `smoothness' condition in \textbf{Prescription 1} means that the components parallel to the $\tau=0$ surface of the tangent of the extremal surface in the two signatures should match at the junction points\footnote{The author thanks K.Narayan for clarifying this prescription.}. According to K. Narayan, this is almost a loaded constraint, but can be understood from the perspective of the replica method. If one constructs a replicated geometry of the Hartle-Hawking state and allows the metric in the bulk fluctuate while keeping the chosen subregion on the holographic screen fixed, as one would do in the usual AdS case \cite{Lewkowycz:2013}, then the saddle which gives the extremal surface will naturally satisfy the `smoothness' condition. For a more detailed description of this point, see section 2 of \cite{GoswamiNarayan:2024}. An explicit derivation by replicated geometry was studied in \cite{Arias:2019pzy}, but the bulk geometry is a Lorentzian de Sitter spacetime with both past infinity and future infinity.

\textbf{Prescription 2} originates from the consideration of the first law of entanglement entropy \cite{Blanco:2013,Bhattacharya:2012}. The first law states that an infinitesimal change in the expectation value of the modular Hamiltonian leads to an infinitesimal change in the entanglement entropy. In the holographic theory, such an infinitesimal change is rephrased as the perturbation of the bulk metric. In the dS/CFT, the density matrix and the modular Hamiltonian in the dual CFT are non-Hermitian, and their expectation values become complex. Nevertheless, we expect that the infinitesimal change in the expectation value of the modular Hamiltonian leads to an infinitesimal change in the pseudo-entropy. The only difference is that the name `first law' is no longer appropriate. In \cite{FujikiKoharaShinmyoSuzuki:2025}, the authors showed that connecting the timelike geodesic in Lorentzian dS to the spacelike geodesic in the Euclidean dS is an obstruction for the `first law' to hold, while extending the integral enables the entanglement pseudo-entropy to satisfy the `first law' in the dS/CFT correspondence.

\textbf{Prescription 3} can be understood by considering the Poincar\'e patch of dS. The metric of the dS Poincar\'e patch can be obtained from that of the AdS Poincar\'e patch by a Wick rotation. The minimal surface in the AdS Poincar\'e patch is spacelike and has a turning point in the bulk, whereas the extremal surface in dS Poincar\'e patch is timelike and has no turning point in the bulk. However, if we complexify the coordinates, the extremal surface in dS Poincar\'e patch can be obtained from that of the AdS Poincar\'e patch by a Wick rotation, hence has a turning point in the complexified bulk spacetime.

The prescription for the extremal surface in de Sitter spacetime remains controversial, because all the known prescriptions give the same results for the pure dS\textsubscript{3} without matter, blackhole or any deformation. For higher dimensions, these prescriptions become problematic by themselves, see \cite{Narayan:2015,Narayan:2022,Narayan:2026,FujikiKoharaShinmyoSuzuki:2025,Doi:2022PRL,Doi:2023JHEP}. Although the extremal surfaces may not exist for generic subregions in higher dimensions, universal properties of pseudo-entropy under infinitesimal deformations of a spherical subregion was studied in \cite{Anastasiou:2025rvz,Anastasiou:2026bbf} using \textbf{Prescription 1}.

In this paper, we focus on \textbf{Prescription 1} and \textbf{Prescription 2} and study them in dS\textsubscript{3}, thus the extremal surfaces are merely geodesics. We aim to study them in a deformed de Sitter spacetime and compare how they would change and whether they would lead to inconsistencies. An interesting deformation is obtained by gluing two de Sitter spacetimes with different radii by a brane. Such a geometry can be naturally generated by considering the Coleman-De Luccia instanton \cite{Coleman-DeLuccia:1980} which was initially proposed for vacuum decay in the presence of gravity. The resulting geometry contains a brane whose tension determined by the potential of the inflaton. For our purpose of dS/CFT, we glue the Euclidean geometry to the Lorentzian geometry at the global time $\tau=0$, in analogy to the Hartle-Hawking no-boundary state \cite{HartleHawking:1983}. One should first solve the Einstein equations to obtain the Euclidean solution and then use it as an input for the Hamiltonian formalism to determine the Lorentzian time evolution. This was worked out in \cite{Cespedes:2020} using the FMP formalism \cite{FMP:1990}. The brane which separates the two de Sitter spacetimes traces out a de Sitter slice, so it is reasonable to expect that the dual CFT is an interface CFT which joins two non-unitary CFTs with different central charges. Moreover, at future infinity, both of de Sitter spacetimes occupy half of the timeslice, no matter how the tension of the brane changes. We note that a similar spacetime was constructed in \cite{Wang:2025jfd} to study the Hilbert space of closed universe.
It is then natural to expect that in such a deformed spacetime, one can see differences between prescriptions for the extremal surface, in the presence of the brane. In particular, the geodesic can end on the brane and probe its conformal data.
Since we don't know how to treat the brane in the complex geometry, we will not study \textbf{Prescription 3} in this paper\footnote{Such a difficulty actually also arises when we apply \textbf{Prescription 2} to our geometry. It is possible for the integral contour to implicitly intersect the brane. We will study this point in detail in section \ref{psudo-entropy}.}.

In the pure de Sitter spacetime, the pseudo-entropy depends only on the size of the subregion and the dS radius, not on its location. But once a brane inserted in the bulk, similar to the case of computing the entanglement entropy of interface CFT by gluing two AdS spacetimes with different radii \cite{Takayanagi:2007,Anous:2022,Wei:2023}, the entanglement entropy behaves differently. Depending on the location of the subregion relative to the interface, the entropy contains a $g$-function which encodes the conformal data of the interface. Since the pseudo-entropy in dS/CFT calculated by any of the three prescriptions matches the analytic continutaion of the entanglement entropy in AdS/CFT, one may expect that this this is also true after the insertion of the brane. However, we will show that such an intuition does not always work. The pseudo-entropy exhibits different behaviors and so does the $g$-function.

In section \ref{Glueing_by_CDL}, we briefly review the Coleman-De Luccia instanton and construct the geometry which glues two de Sitter spacetimes with different radii in general dimensions. 
In section \ref{dS_ICFT}, we argue that it is reasonable for the geometry at hand to have an interface CFT joining two non-unitary CFTs with different central charges as its holographic dual, provided that the dS/CFT correspondence is exact. We give the dS/ICFT dictionary as an analog of the dS/CFT dictionary.
In section \ref{psudo-entropy}, we compare \textbf{Prescription 1} and \textbf{Prescription 2} for the holographic computation of entanglement entropy. We argue that the `smoothness' constraint, in \textbf{Prescription 1}, imposed at the junction points of timelike and spacelike geodesics will lead to an inconsistency with general properties of interface CFTs. After pointing out the inconsistency of \textbf{Prescription 1}, we adopt \textbf{Prescription 2} for the remainder of this paper. We find that \textbf{Prescription 2} shares some similarities with the AdS/ICFT case \cite{Takayanagi:2007,Anous:2022,Wei:2023}, while also exhibiting differences. We also obtain the $g$-function by computing the pseudo-entropy associated with a subregion perpendicular to the interface.
In section \ref{conclusion}, we conclude our results and comment on future work.

\section{The de Sitter-de Sitter decay}\label{Glueing_by_CDL}
In this section, we construct a geometry which glues two de Sitter spacetimes with different radii. Such a geometry can be naturally generated by considering the Coleman-De Luccia instanton \cite{Coleman-DeLuccia:1980} which was initially proposed for vacuum decay in the presence of gravity.
It starts from introducing an inflaton whose potential has two positive local minima.
Applying a symmetric ansatz and the thin-wall approximation, the inflaton takes kink-like solution. After absorbing the region where the inflaton changes dramatically by a brane, the resulting geometry consists of two de Sitter spacetimes with different radii separated by a brane whose tension is determined by the potential. The Euclidean solution can be geometrically regarded as a spacetime obtained by gluing two spheres with different radii at a given polar angle.
The trajectory of the brane is obtained after changing to the Lorentzian signature \cite{Cespedes:2020}.
Since there is no obstruction to the generalization to general dimensions, we will work in $d$-dimension throughout this section.

\subsection{Description for dS-dS decay}
We briefly review the {\it Coleman-De Luccia (CDL) instanton}, using the de Sitter-de Sitter decay as an example. We restrict ourselves to the regime where the {\it thin-wall approximation} applies. For an analysis of CDL instanton beyond the thin-wall approximation, see \cite{XiDong:2011}\footnote{We should emphasize that the Lorentzian geometry in \cite{XiDong:2011} is obtained by analytic continuation of coordinates and does not render a satisfactory geometry for the purpose of studying holography of the combination of two de Sitter spacetimes. We will discuss this point under thin-wall approxination in the next subsection.}.

\begin{figure}[h]
    \centering
    \includegraphics[width=4cm]{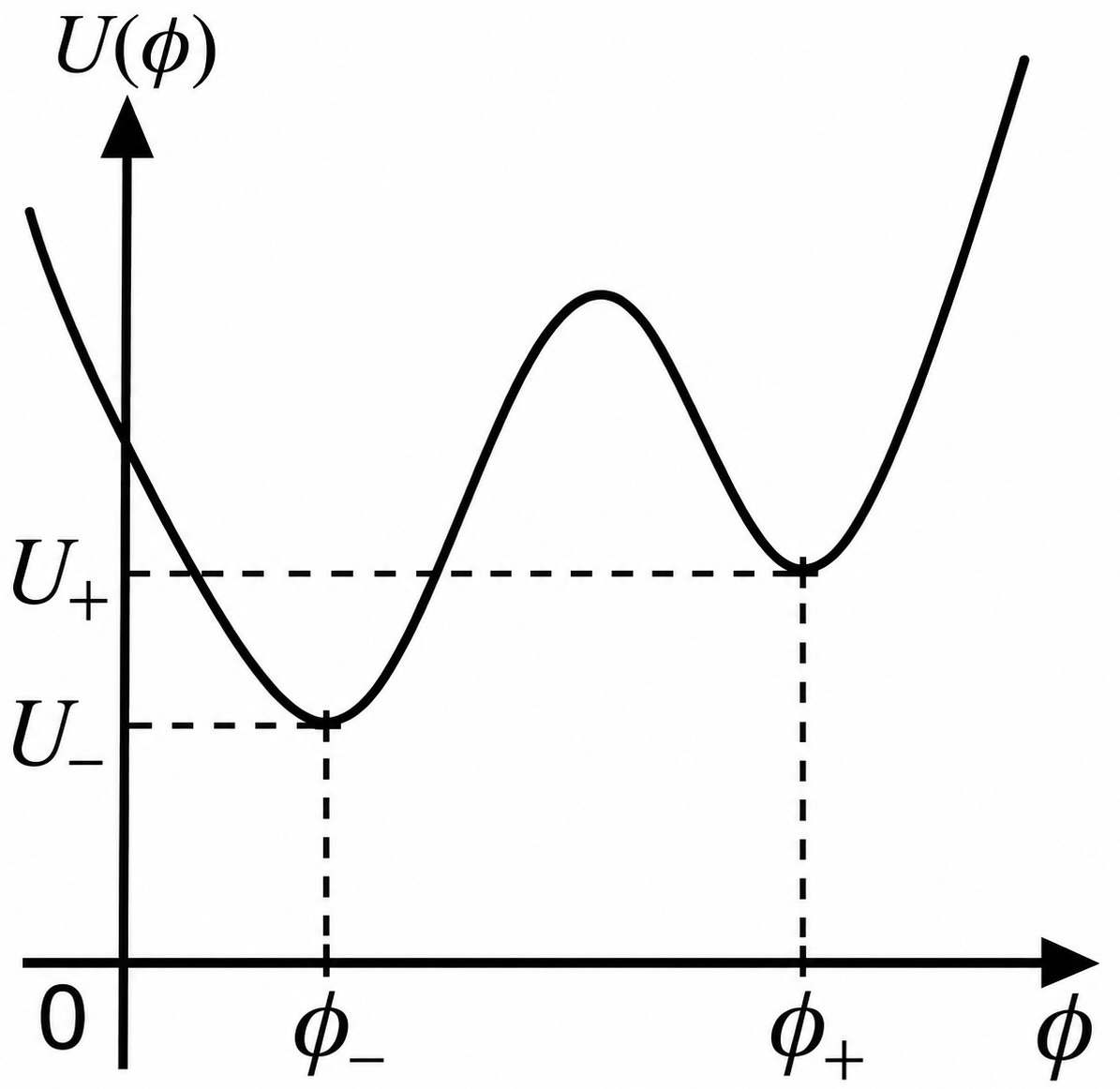}
    \caption{The potential of inflaton $\phi$, which is positive definite and has two positive local minimum. $\phi=\phi_+$ is the fake vacuum and $\phi=\phi_-$ is the true vacuum. There is a barrier between the two local minimum which is maximized at $\phi=\bar{\phi}$.}
    \label{potential}
\end{figure}

Consider the $d$-dimensional Einstein-Hilbert action with an inflaton $\phi$ in Euclidean signature,
\begin{equation}\label{action}
    S_E=-\frac{1}{16\pi G}\int d^dx\sqrt{g}R+\int d^dx\sqrt{g}\left(\frac{1}{2}\nabla_\mu\phi\nabla^\mu\phi+U(\phi)\right)\,.
\end{equation}
The value of the potential $U(\phi)$ serves as the cosmological constant. Since we aim to study the dS-dS decay, the potential is supposed to take the shape shown in Fig.\ref{potential}. The states where $\phi=\phi_+$ and $\phi=\phi_-$ are respectively the fake vacuum and the true vacuum. As in the quantum field theory, one expects that the fake vacuum will decay into the true vacuum \cite{Coleman-DeLuccia:1980}. In the spacetime, one would observe the nucleation of a bubble of true vacuum within the fake vacuum. The Coleman-De Luccia instanton is a nontrivial solution to the equations of motion whose $\phi$ approaches $\phi_+$ at the asymptotic region and $\phi_-$ at the center of the bubble which minimizes the action.

The original analysis by Coleman and De Luccia \cite{Coleman-DeLuccia:1980} was based on the assumption that the solution minimizing the action would be $O(d)$-symmetric. This conjectured symmetry has been confirmed in the anti de Sitter case in \cite{Oshita:2023} using holographic methods, but remains open in general. Although there is no proof of the $O(d)$ symmetry in the dS-dS decay, no counter-example has been constructed, nor is there any evidence for the existence of such counter-examples so far. Therefore, we will also adopt the $O(d)$ symmetric ansatz in the following. The metric and the inflaton then is only described by the radial coordinate $\xi$,
\begin{equation}
    ds^2=d\xi^2+f(\xi)^2\,d\Omega^2_{d-1} \,,\quad
    \phi=\phi(\xi)
\end{equation}
where $d\Omega^2_{d-1}$ denotes the metric of the $d-1$-dimensional unit sphere $S^{d-1}$. 
We choose the center of the bubble to be $\xi=0$ and denote the radius of the bubble measured from the center by $R_0=f(\bar{\xi})$.
The Euclidean action then reduces to
\begin{equation}
    S_E=\text{Vol}(S^{d-1})\int d\xi\left[f^{d-1}\left(\frac{1}{2}(\phi')^2+U(\phi)\right)-\frac{d-1}{16\pi G}\left(-2f''f^{d-2}+(d-2)f^{d-3}(1-f'^2)\right)\right]
\end{equation}

The equation of motion associated with $\phi$ and the nontrivial component of the Einstein equations are given by 
\begin{align}
    &f''+(d-1)\frac{f'}{f}\phi'=\frac{\partial U}{\partial\phi} \label{inflaton_EOM}\\
    &(f')^2=1+\frac{16\pi G}{(d-1)(d-2)}f^2\left(\frac{1}{2}(\phi')^2-U(\phi)\right) \label{Einstein_eq}
\end{align}
where $'$ denotes the derivative with respect to $\xi$.
The thin-wall approximation says that the second term in equation (\ref{inflaton_EOM}) is negligible throughout the domain of $\xi$, that is, $\phi$ is nearly constant both inside and outside the bubble while the region where $\phi$ interpolates between $\phi_+$ and $\phi_-$ is sufficiently small. Under this approximation, equation (\ref{inflaton_EOM}) can be written as
\begin{equation}
    \frac{1}{2}(\phi')^2-U(\phi)=-U(\phi_\pm)
\end{equation}
which implies that $\phi$ goes monotonically from $\phi_-$ to $\phi_+$ as $\xi$ increases. The solution of $\phi$ can then be expressed as
\begin{equation}
    \int_{\bar{\phi}}^\phi d\phi\left[2(U(\phi)-U(\phi_\pm))\right]^{-1/2}=\xi-\bar{\xi}
\end{equation}
where $\bar{\phi}=\phi(\bar{\xi})$ maximizes the barrier shown in Fig.\ref{potential} and satisfies $\phi_-<\bar{\phi}<\phi_+$.
The constant $U(\phi_\pm)$ depends on which side of the bubble we are considering. 
The radius of the bubble $R_0$ is found by extremizing the decay rate $\Gamma\propto\,e^{-B/\hbar}(1+O(\hbar))$\footnote{The Planck constant $\hbar$ is kept explicit only here and left implicit everywhere else.} which is equivalent to extremizing the exponent
\begin{align}
    B&=S_E[\phi]-S_E[\phi_+] \notag\\
    &=\text{Vol}(S^{d-1})\int d\xi\left[2f^{d-1}U(\phi)-\frac{(d-1)(d-2)}{8\pi G}f^{d-3}\right]
    -(\phi\to\phi_-)
\end{align}
where $\phi$ on the RHS denotes the Coleman-De Luccia instanton.
Under the thin-wall approximation, $\phi$ is approximated by $\phi_-$ (or $\phi_+$) inside (or outside) the wall, thus the exponent can be evaluated piecewisely as
\begin{equation}
    B=B_{\text{outside}}+B_{\text{wall}}+B_{\text{inside}}
\end{equation}
with each term given by
\begin{align}
    &B_{\text{outside}}=0  \\
    &B_{\text{wall}}=\text{Vol}(S^{d-1})\,R_0^{d-1}\,\sigma \\
    &B_{\text{inside}}=-\frac{(d-1)(d-2)}{8\pi G}\text{Vol}(S^{d-1})\int_0^{R_0}df f^{d-3}\left(1-\frac{16\pi G\,U_-}{(d-1)(d-2)}f^2\right)^{1/2}  \notag\\
    &\hspace{45pt}+(U_-\to U_+)
\end{align}
where $\sigma$ can be considered as the tension of the wall,
\begin{equation}
    \sigma=2\int_{R_0-\delta f}^{R_0+\delta f}
    (U(\phi)-U(\phi_+))\,.
\end{equation} 
As can be seen from the expression, the tension $\sigma$ depends on the potential $U(\phi)$ and can be tuned by changing the height and shape of the potential barrier while keeping the values of $U_\pm$ fixed. Thus, in the following, we will treat the tension $\sigma$ as a free parameter. Solving the equation $\partial B/\partial R_0=0$, we find that the radius of the bubble is given by
\begin{equation}
    R_0=\frac{(d-1)(d-2)\sigma}{\sqrt{(d-2)^2(U_+-U_-)^2+8(d-1)(d-2)G\pi \sigma^2(U_++U_-)+16(d-1)^2G^2\pi^2\sigma^4}}\,.
\end{equation}
Now we turn to solving the Einstein equation (\ref{Einstein_eq}) both inside and outside the bubble. Again, under the thin-wall approximation, the $\phi'^2$ term can be neglected and $U(\phi)$ is approximated by $U_\pm$,
\begin{equation}
    (f')^2=1-\frac{16\pi G}{(d-1)(d-2)}U_\pm \,.
\end{equation}
The solution is 
\begin{equation}
    f(\xi)=\sqrt{\frac{(d-1)(d-2)}{16\pi G\,U_\pm}}\sin\left[\sqrt{\frac{16\pi G\, U_\pm}{(d-1)(d-2)}}(\xi-\xi_0)\right]
\end{equation}
which describes a Euclidean dS\textsubscript{d} spacetime with dS radius 
\begin{equation}\label{dS_radius}
    l_\pm=\sqrt{\frac{(d-1)(d-2)}{16\pi G\,U_\pm}}\,.
\end{equation}
Since $U_-< U_+$, the true vacuum has a larger dS radius ($l_-$) than the fake vacuum ($l_+$). In terms of the dS radius (\ref{dS_radius}), the radius of the bubble can be written as 
\begin{equation}
    R_0^2=\frac{4 \mathcal{E}^2}{(l_+^{-2}-l_-^{-2})^2+2(l_+^{-2}+l_-^{-2})\mathcal{E}^2+\mathcal{E}^4}
\end{equation} 
where we have introduced $\mathcal{E}=\frac{8\pi G}{d-2}\sigma$ which represents the energy density of the brane. One can easily verify that $R_0$ is smaller than $l_\pm$ for arbitrary $\mathcal{E}$.
Now the physical picture is clear. Under the thin-wall approximation, $\phi$ takes $\phi_-$ inside the bubble and $\phi_+$ outside the bubble, and the potential $U(\phi)$ taking values $U_-$ inside and $U_+$ outside the bubble serves as the cosmological constant in the action (\ref{action}). On the other hand, the region in which $\phi$ interpolates from $\phi_-$ to $\phi_+$ is sufficiently small and together with an integral over the potential can be absorbed into a brane with tension $\sigma$. Therefore, approximately there is no dynamics in the inflaton on the two sides of the brane, where the spacetime can be treated as pure de Sitter spacetime. The spacetime looks like portions of two spheres with radii $l_\pm$ being glued at radius $R_0$, see Fig.\ref{Gluing_sphere}.

\subsection{Trajectory of the thin-wall}\label{sec:trajectory}
Due to the unusual metric describing the gluing of two Euclidean de Sitter spacetimes, the Lorentzian spacetime obtained by analytic continuation of coordinates is not suitable. For our purpose of de Sitter holography, we need the Lorentzian spacetime to be a bubble expanding in the global de Sitter spacetime. However, naive analytic continuation will give an open universe which only covers a portion of the global de Sitter spacetime.
This is most clear from the embedding equations in $\mathbb{R}^{d+1}$, 
\begin{equation}\label{embedding}
    X_0=\sin\xi\sin\theta, \, 
    X_i=\sin\xi\cos\theta\,\Omega_i\,(i=1,\ldots,d-1),\,
    X_d=-\cos\xi
\end{equation}
where $\sum_i\Omega_i^2=1,\,\xi\in[0,\pi]$, and we have set the dS radius to be 1 for clarity. Then the induced metric is identical to the solution obtained in the last subsection,
\begin{equation}\label{induced_metirc}
    ds^2=d\xi^2+\sin^2\xi(d\theta^2+\cos\theta^2d\Omega^2_{d-2})\,.
\end{equation}
The analytic continuation of $\xi$ introduces more than one timelike coordinates, whereas in the conventional global coordinates it only introduces one. One may also try to analytically continue $\theta$, but this will leave $X_d$ still compact and does not reproduce the global de Sitter spacetime either. The difficulty originates from the fact that the embedding (\ref{embedding}) used here employs a different foliation from that of usual global coordinates.

Instead, we need to first perform the Euclidean path integral on the lower half of the deformed sphere to prepare what we call {\it Coleman-De Luccia state} $\Psi_{\text{CDL}}$, as an analog of Hartle-Hawking state, and then act on this state with the time evolution operator $e^{it H}$ to find its time evolution in Lorentzian signature. The Lorentzian time evolution then follows by working out Hamiltonian formalism. In particular, the resulting Lorentzian time evolution of the Coleman-De Luccia state at hand was worked out in \cite{Cespedes:2020} using the so-called FMP formalism  developed in \cite{FMP:1990}.

In terms of the proper time on the brane $t$, the trajectory is given by \cite{Cespedes:2020}
\begin{equation}
    R(t)=R_0\cosh\frac{t}{R_0}
\end{equation}
which is the same as the trajectory of a particle in the Minkowski spacetime being accelerated at $t=0$ with zero velocity and constant acceleration $1/R_0$. This is easy to understand, because the cosmological constants are different on the two sides of the brane and they generate a force on the brane which pushes it outward the center.

\begin{figure}[h]
  \begin{minipage}[b]{0.5\columnwidth}
    \centering
    \includegraphics[width=6cm]{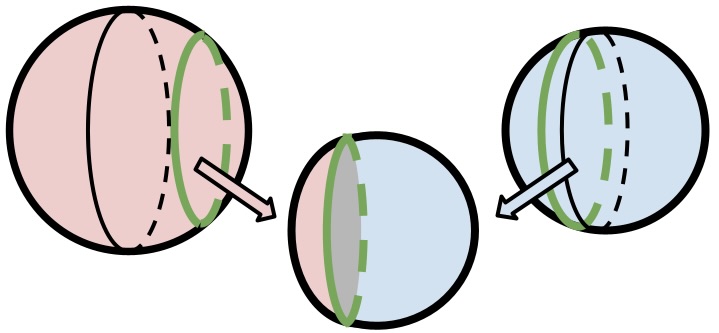}
    \caption{Gluing two spheres with different radii. The blue sphere has a smaller radius while the red one has a larger radius. A samll portion from the red sphere is glued to a large portion of the blue one at radius $R_0$.}
    \label{Gluing_sphere}
  \end{minipage}
  \hspace{0.04\columnwidth} 
  \begin{minipage}[b]{0.47\columnwidth}
    \centering
    \includegraphics[width=3cm]{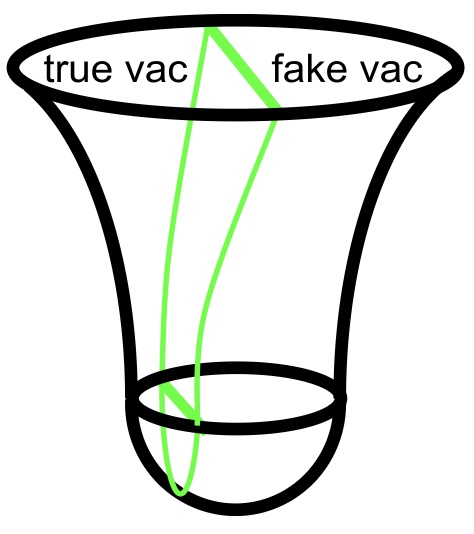}
    \caption{Trajectory of the brane in the global de Sitter spacetime. The brane is drawn in green. The upper half is in Lorentzian signature and the lower half is in Euclidean signature. }
    \label{CDL_state}
  \end{minipage}
\end{figure}

Then we need to embed the trajectory into the de Sitter spacetime. Although we have used a unconventional metric (\ref{induced_metirc}) for the Euclidean signature, to illustrate the trajectory of the brane, it is most convenient to use the conformal coordinates as in \cite{Cespedes:2020}:
\begin{equation}\label{global_dS}
    ds^2=\frac{l_\pm^2}{\cos^2T_\pm}\left(-dT^2_\pm+d\rho^2_\pm+\sin^2\rho_\pm\,d\Omega^2_{d-2}\right)
\end{equation}
where $-$ denotes the coordinates for the true vacuum and $+$ for the fake vacuum. The trajectory of the brane in the global de Sitter spacetime is then 
\begin{equation}\label{trajectory_of_brane}
    \cos\rho_\pm=\sqrt{1-\frac{R_0^2}{l_\pm^2}}\,\cos T_\pm
\end{equation}
which is drawn in Fig.\ref{CDL_state}.
We first describe the geometry of the resulting spacetime here. Each time slice of the global dS\textsubscript{d} (\ref{global_dS}) is a sphere $S^{d-1}$ whose radial direction is parametrized by $\rho$. Since the Coleman-De Luccia state is used as an initial condition, the induced metric on the $T=0$ slice of the Lorentzian spacetime (\ref{global_dS}) and should match that of the Euclidean spacetime (\ref{induced_metirc}). This means that the center of the true and fake vacuum which respectively correspond to $\xi=0,\pi$ in the Euclidean signature should correspond to $\rho=0,\pi$ in the Lorentzian signature, which determines brane trajectory to be equation (\ref{trajectory_of_brane}).
More explicitly, the metric is written piecewisely as 
\begin{equation}\label{Lorentzian_metric}
    ds^2=\left\{
    \begin{aligned}
        &\frac{l_-^2}{\cos^2T_-}(-dT_-^2+d\rho_-^2+\sin^2\rho_-\,d\Omega_{d-2}^2)\,,
        & \rho_-<\cos^{-1}\left(\sqrt{1-\frac{R_0^2}{l_-^2}}\cos T_-\right)\\
        &\frac{l_+^2}{\cos^2T_+}(-dT_+^2+d\rho_+^2+\sin^2\rho_+\,d\Omega_{d-2}^2)\,,
        & \rho_+>\cos^{-1}\left(\sqrt{1-\frac{R_0^2}{l_+^2}}\cos T_+\right)
    \end{aligned}
    \right.
\end{equation} 

We also note that at global time $T=0$, the location of the brane in the $\pm$ coordinates is given by 
\begin{equation}
    l_\pm\,\sin\rho_\pm=R_0 \,.
\end{equation}
The inequalities $R_0<l_\pm$ and $l_+<l_-$ imply that neither the portion taken from the sphere with a larger radius nor the portion from the sphere with a smaller radius can exceed a hemisphere. As time evolves, the brane will expand and the true vacuum will gradually occupies an increasing fraction of each time slice, as depicted in Fig.\ref{CDL_state}. Interestingly, the brane will cease to expand at future infinity $T=\pi/2$ and the true vacuum will eventually occupy exactly half of the time slice.

The final property of the resulting spacetime that we need to mention here is the symmetry which is relavent when we consider its holographic dual in the next section. The region inside and outside the brane are both locally de Sitter but with different dS radius, so they locally have the same symmetry $SO(1,d)$ as the usual de Sitter spacetime. On the other hand, as already noticed in \cite{Cespedes:2020}, the brane traces out a dS\textsubscript{d-1} slice of the global dS\textsubscript{d} and has the symmetry $SO(1,d-1)$. To see this, we need the induced metirc  
\begin{equation}
    ds^2=-\frac{R_0^2}{\cos^2T(\sin^2T+R_0^2\cos^2T/l^2)}dT^2 + (l^2\tan^2T+R_0^2)d\Omega_{d-2}^2 \,.
\end{equation}
After the coordinate change
\begin{equation}\label{eq:coordinate_change}
    \sinh\frac{\tau}{R_0}=\frac{l}{R_0}\tan T\,,
\end{equation}
the induced metric becomes
\begin{equation}
    ds^2=-d\tau^2+R_0^2\cosh^2\frac{\tau}{R_0}\,d\Omega_{d-2}^2
\end{equation}
which is identical to the metric of dS\textsubscript{d-1} with radius $R_0$.
From equation (\ref{eq:coordinate_change}), the first Israel conjunction condition tells us that the coordinates $T_-$ and $T_+$ are related by
\begin{equation}
    l_- \tan T_- = l_+ \tan T_+ \,.
\end{equation}
at the brane.

Unlike the AdS foliation of AdS spacetime which fills in the entire spacetime, the dS foliation of dS only covers a portion of the full spacetime. Thus, no matter how the radius $R_0$ changes, the brane is restricted in the region that is foliated by dS slicing. 

The properties of the spacetime obtained by the Coleman-De Luccia instanton discussed here will play an important role in the holographic dual and the holographic computation of entanglement (pseudo-)entropy in the following sections.

\section{The dS/ICFT correspondence}\label{dS_ICFT}
In this section, we argue that it is reasonable to expect the spacetime constructed in the previous section to have an interface CFT (ICFT) as its holographic dual. We first review how the dS/CFT correspondence \cite{Strominger:2001,Maldacena:2002} works. Based on the hypothetical dS/CFT correspodence, we argue that the dual theory to our spacetime consists of two non-unitary CFTs with different central charges joined by an interface. 

\subsection{General description for dS/CFT correspondence}
The dS/CFT correspondence \cite{Strominger:2001,Maldacena:2002} suggests that quantum gravity in the $d+1$ dimensional asymptotically de Sitter spacetime is dual to the $d$ dimensional conformal field theory living at future infinity. As an analog of the GKPW dictionary of the AdS/CFT correspodence \cite{GKP:1998,Witten:1998}, there is a proposed dictionary for the dS/CFT. Here we cite from \cite{Doi:2023JHEP},
\begin{equation}\label{dS/CFT_dictionary}
    Z_{\text{CFT}}[\phi_0,\gamma]=\underset{g\to\gamma}{\int_{\phi\to\phi_0}}[d\phi\,dg]\,e^{iI_{\text{dS}}[\phi,g]}\Psi_{\text{HH}}
\end{equation}
where $\Psi_{\text{HH}}$ denotes the Hartle-Hawking state \cite{HartleHawking:1983} prepared by the Euclidean path integral over a hemisphere. The $\phi$ and $g$ denotes the matter field and the metric in the bulk while $\phi_0$ and $\gamma$ denotes the corresponding boundary condition on the future infinity. Given the Hartle-Hakwing state (a hemisphere) as an input, the path integral in the Lorentzian signature is dominated by the global de Sitter spacetime.

The relation between the scaling dimension $\Delta$ of a boundary opeartor and the mass of its corresponding scalar field in the bulk is given by \cite{Strominger:2001}
\begin{equation}
    m^2l^2_{\text{dS}}=\Delta(d-\Delta)
\end{equation}
or equivalently
\begin{equation}
    \Delta_\pm=\frac{d}{2}\pm\frac{1}{2}\sqrt{d^2-4m^2l^2_{\text{dS}}} \,.
\end{equation}
We see that if a scalar in the bulk spacetime has mass larger than $d^2/(4l_{\text{dS}}^2)$, the corresponding operator has a complex scaling dimension. The conformal symmetry at the asymptotic boundary generated by the Killing vectors has a real central charge $\tilde{c}$, whereas the dual CFT\textsubscript{d} is known to have a complex central charge \cite{Maldacena:2002},
\begin{equation}
    c=(-i)^{d-1} \tilde{c}\,.
\end{equation}
These features are the signs of a non-unitary CFT whose density matrices and modular Hamiltonians are in general non-Hermitian. Thus the entanglement entropy is expected to be complex and should be recognized as the entanglement pseudo-entropy \cite{Nakata:2020}, which was confirmed in a concrete construction for the dS\textsubscript{3}/CFT\textsubscript{2} in \cite{Hikida:2022}. In particular, in the dS\textsubscript{3}/CFT\textsubscript{2} case, the central charge of the dual CFT is 
\begin{equation}
    c=-i\tilde{c}\,,\quad 
    \tilde{c}=\frac{3l_{\text{dS}}}{2G}\,.
\end{equation}

\subsection{The dual CFT for dS-dS decay}
As introduced in section\ref{sec:trajectory}, the Euclidean path integral over the deformed hemisphere obtained by gluing two spheres with different radii prepares what we call the Coleman-De Luccia state $\Psi_{\text{CDL}}$. The dictionary for the dS/CFT (\ref{dS/CFT_dictionary}) then replaces $\Psi_{\text{HH}}$ by $\Psi_{\text{CDL}}$ and the action $I_{\text{dS}}$ by $I_{\text{CDL}}$ which includes the usual Einstein-Hilbert action, the Gibbons-Hawking term, the cosmological constant together with the step function and a matter term associated with the brane tension, see equation (5.3) of \cite{Cespedes:2020}. In short, the dictionary (\ref{dS/CFT_dictionary}) is modified as follows:
\begin{equation}\label{dS/ICFT_dictionary}
    Z_{\text{CFT}}[\phi_0,\gamma]=\underset{g\to\gamma}{\int_{\phi\to\phi_0}}[d\phi\,dg]\,e^{iI_{\text{CDL}}[\phi,g]}\Psi_{\text{CDL}}
\end{equation}
where the measure $dg$ also includes the fluctuations of the brane trajectory.
Since the spacetime obtained in section\ref{sec:trajectory} satisfies the Einstein equations and the Israel junction conditions, it dominates the path integral on the right-hand side of equation (\ref{dS/ICFT_dictionary}) in the semiclassical limit $G\to0$.

The true vacuum and the fake vacuum have dS radii $l_-$ and $l_+$ respectively, whose respective CFTs have central charges of different magnitudes. Thus, the quantum gravity in the two regions separated by the brane is supposed to be encoded in two different CFTs at future infinity. As was explored in section\ref{sec:trajectory}, the brane joining the two regions has traces out a dS slice and thus preserves part of the symmetry of the global dS. It is supposed to be dual to an interface joining the two CFTs lying on the future infinity.

In this way, we have established the correspondence between the de Sitter spacetime and the non-unitary interface CFT, the dS/ICFT correspondence, based on the hypothetical dS/CFT correspondence. In the next section, we will use this correspondence to test the first two prescriptions for the holographic computation of the entanglement entropy in dS/CFT.

\section{Pseudo-entropy in the dS/ICFT}\label{psudo-entropy}
In this section, we use the dS/ICFT correspondence established in the previous section as a laboratory to test the prescriptions for the holographic computation of pseudo-entropy in the dS/CFT. We will start by reviewing the issues encountered in the prescription of the extremal surface. 
We focus on dS\textsubscript{3}/ICFT\textsubscript{2}, where the extremal surfaces are simply geodesics. We will choose the subregion to be centered at the center of the true vacuum, at the center of the fake vacuum and at the interface. Surprisingly, we will see that certain prescriptions will cause inconsistency with universal properties of interface CFTs, and that the results do not always agree with the analytically continued results in the AdS/ICFT \cite{Takayanagi:2007,Anous:2022,Wei:2023}. Nevertheless, since both the bulk spacetimes in AdS/ICFT and dS/iCFT shares a similarity that in both cases the spacetime with a smaller radius occupies a larger portion \cite{Wei:2023}, which is relevant when determining whether the geodesics would intersect the brane,  we still consider the analytic continuation of the results in the AdS/ICFT as a guide. When the results do match after analytic continuation, it is a strong evidence that the prescription under analysis is consistent.

\subsection{Issues of holographic computation of entropy}
As in the AdS/CFT correspondence, where the entanglement entropy associated with a subregion $A$ in the boundary CFT is computed by the area of the extremal surface anchored on the boundary of $A$ \cite{RT:2006,HRT:2007}, in dS/CFT one also expects that the entanglement entropy of a subregion in the dual CFT is similarly computed by the area of the extremal surface whose boundary coincides with that of the chosen subregion. However, as first noticed in \cite{Narayan:2015}, the geodesics in dS\textsubscript{3} are timelike and do not close. A more serious issue is that in higher dimensions the extremal surface does not always exist for generic subregions \cite{Narayan:2015,Narayan:2017,Narayan:2022,Narayan:2026,FujikiKoharaShinmyoSuzuki:2025,Doi:2023JHEP}. Only for certain shapes of the subregion does the extremal surface exist, such as the hemisphere at future infinity.

As discussed in the introduction, there are several different prescriptions for extremal surfaces all of which give the same result in pure dS\textsubscript{3} spacetime\footnote{Here by pure dS we mean the de Sitter spacetime without any insertions of branes or matter.}. For clarity, consider an interval at future infinity as $\rho_1\leq\rho\leq\rho_2$ and $\phi=0$ where $0\leq\rho_1<\rho_2\leq\pi$ and $\phi$ is the azimuthal coordinate. If we impose only the boundary condition that $\rho=\rho_1$ at $T=\pi/2$, the geodesic length will be minimized at a straight line down to $T=0$. Only after adding the additional condition that when $\rho=\rho_2$ the global time $T$ also takes $\pi/2$, do we get the correct geodesic:
\begin{equation}
    \sin\left(\rho\pm\frac{\pi}{2}-\frac{\rho_1+\rho_2}{2}\right)=\cos\frac{\rho_1-\rho_2}{2}\,\sin T \,.
\end{equation}
The equation consists of two disconnected geodesics. One starts from $\rho=\rho_1,\,T=\frac{\pi}{2}$ and ends at $\rho=\frac{\rho_1+\rho_2}{2}-\frac{\pi}{2},\,T=0$ while the other starts from $\rho=\rho_2,\,T=\frac{\pi}{2}$ and ends at $\rho=\frac{\rho_1+\rho_2}{2}+\frac{\pi}{2},\,T=0$. The two endpoints on the $T=0$ slice shift in the $\rho$-direction by $\frac{\pi}{2}$, though in the opposite direction, from the center of the interval and are therefore antipodal. The total geodesic length is then calculated as 

\begin{equation}
    \text{Area}(\gamma)=2i l_{\text{dS}}\log
    \left(\frac{2}{\epsilon}\sin\frac{\rho_2-\rho_1}{2}\right)
\end{equation}
where $\epsilon=2e^{-\tau_\infty}$ is the UV cutoff in terms of the global coordinates $\tau=\sinh^{-1}(\tan T)$.
This is not satisfactory yet since the timelike geodesic does not close.
Here comes the \textbf{Prescription 1} \cite{Narayan:2022} which says that one should smoothly connect the timelike geodesic in the Lorentzian dS to a geodesic connecting the two endpoints at $T=0$ in the Euclidean dS. In this way, one makes the geodesic closed. Thanks to the fact that the two endpoints are antipodal, it is possible for the geodesic to extend into the interior of the hemisphere \cite{Narayan:2022,Doi:2023JHEP,Doi:2022PRL}. Otherwise, one can only connect the two points by a geodesic lying along the rim.
On the other hand, \textbf{Prescription 2} cite{FujikiKoharaShinmyoSuzuki:2025} suggests that when calculating the geodesic length as an integral over $\rho$ one should analytically continue the coordinate $\rho$ to the imaginary direction. After the analytic continuation of the integrand, the geodesic closes in the complex $\rho$-plane at $i\infty$.
\textbf{Prescription 3} \cite{Narayan:2026} says that since the dS spacetime can be obtained from the AdS spacetime after the analytic continuation $l_{\text{AdS}}\to il_{\text{dS}}$, one can consider the analytic continuation of the Poincar\'e patch together with the geodesic. The geodesic length will simply be the analytic continuation of that of the AdS\textsubscript{3} spacetime.

All the descriptions described above agree with the total geodesic length:
\begin{equation}
    \text{Area}(\gamma)=\pi l_{\text{dS}} + 2il_{\text{dS}}\log
    \left(\frac{2}{\epsilon}\sin\frac{\rho_2-\rho_1}{2}\right)\,.
\end{equation}
According to the RT formula \cite{RT:2006}, the entanglement entropy of the chosen interval is then calculated as 
\begin{equation}
    S_A=\frac{\text{Area}(\gamma)}{4G}
    =i\frac{\tilde{c}}{3}\log\left(\frac{2}{\epsilon}\sin\frac{\rho_2-\rho_1}{2}\right)+\frac{\pi\tilde{c}}{6}
\end{equation}
where $\tilde{c}=3l_{\text{dS}}/2G$ \footnote{In following sections, we denote the dS radius only by $l$.}. 
In the rest of this section, we will test \textbf{Prescription 1} and \textbf{Prescription 2} for extremal surfaces by applying them to various subregions in the dS/ICFT. The reason we do not study the \textbf{Prescription 3} is that it is not clear how to treat the brane in the complexified geometry. Thus, we leave the study of \textbf{Prescription 3} for a future work.

\subsection{Subregion centered at fake vaccum and an inconsistency}
As a first test, consider an interval $A$ of size $2\theta$ centered at the center of the fake vacuum region:
\begin{equation}\label{eq:subregion1}
    A=\{\pi-\theta\leq \rho_+\leq\pi,\,\phi=0\}\cup
    \{\pi-\theta\leq \rho_+\leq\pi,\,\phi=\pi\}
\end{equation}
where $\theta\in[0,\frac{\pi}{2}]$.

\begin{figure}[h]
  \begin{minipage}[b]{0.5\columnwidth}
    \centering
    \includegraphics[width=3.7cm]{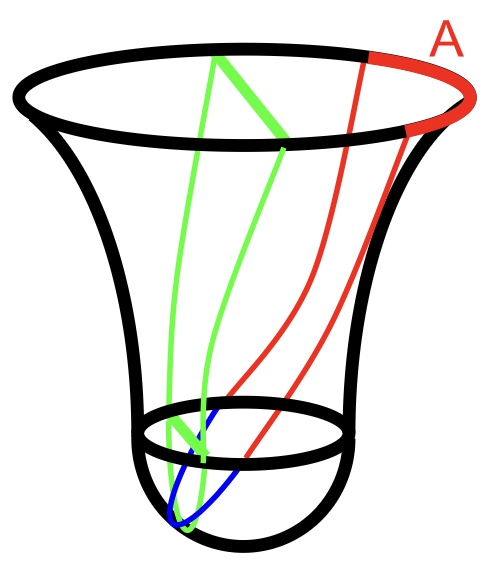}
    \caption{Subregion centered at the center of the fake vacuum. The timelike geodesic and spacelike geodesic are respectively shown in red and blue.}
    \label{Fig:A_fake_vac}
  \end{minipage}
  \hspace{0.04\columnwidth} 
  \begin{minipage}[b]{0.47\columnwidth}
    \centering
    \includegraphics[width=5cm]{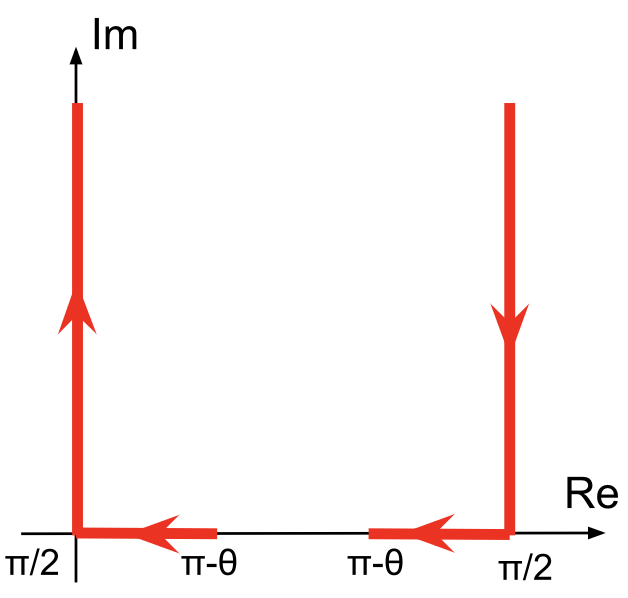}
    \caption{The integral contour of $\rho$ closes at $i\infty$. The geodesic and the brane do not have any complex intersection points.}
    \label{Fig:integral_contour1}
  \end{minipage}
\end{figure}

We start our analysis by applying \textbf{Prescription 1} for extremal surfaces \cite{Narayan:2022}. In this case, the two disconnected timelike geodesics start from $\rho_+=\pi-\theta$ at future infinity, descend to the point $\rho_+=\pi/2$ at the global time $T_+=0$, as shown in Fig.\ref{Fig:A_fake_vac}. The spacelike geodesic is then an arc connecting the two endpoints at $T_+=0$.
For the maximal subregion, $\theta=\frac{\pi}{2}$, the geodesics remain vertical while extending downward and never intersect the brane, just as in the pure dS case \cite{Narayan:2015,Narayan:2017,Narayan:2022,Narayan:2026,Doi:2023JHEP,Doi:2022PRL}.
The entanglement entropy of $A$ computed by the length of the geodesic simply reads
\begin{equation}
    S_A=i\frac{\tilde{c}_+}{3}\log\frac{2}{\epsilon}+\frac{\pi\tilde{c}_+}{6}
\end{equation}
where $\tilde{c}_+=3l_+/2G$.
At first sight, one may think that the entanglement entropy associated with the maximal subregion does not probe any information about the interface or the CFT on the other side is very surprising, since on the CFT side $S_A$ is calculated by the two-point function of twist operators which is supposed to be affected by the interface when the insertions are close to it. 
To understand this, we need to recall the case of AdS/ICFT \cite{Anous:2022,Wei:2023}. To construct the bulk spacetime dual to an interface CFT, one needs to glue two AdS spacetimes with different radii along a brane which bends toward the side with a larger radius. The geodesic connecting two points on the boundary of the smaller radius never intersects the brane and thus does not probe any information of the interface. The same reasoning applies to our case of dS/ICFT, since in both cases the spacetime with a smaller radius occupies a larger portion \cite{Wei:2023}. One important difference, however, is that in our case if $A$ is not centered at $\rho=\pi$, it is possible for the timelike geodesic to intersect the brane and thus probe the information of the interface.

Next, let us illustrate an inconsistency of the \textbf{Prescription 1} that arises as $\theta$ decreases. As $\theta$ decreases and the size of the subregion $A$ shrinks, the timelike geodesics still end at $\rho_+=\pi/2$ at $T_+=0$ but the momentum along the $T=0$ slice at that point is no longer zero, that is, the geodesics will tilt. Then the `smoothness' condition \cite{Narayan:2022,Narayan:2026} bends the spacelike geodesic toward the brane. Once $\theta$ decreases below a critical value $\theta_c$, the spacelike geodesic will intersect the brane, as shown in Fig.\ref{Fig:A_fake_vac} and the real part of entanglement entropy $S_A$ will probe the information of the interface. 
This is inconsistent when we consider the computation of entanglement entropy on the CFT side. As mentioned above, on the CFT side the entanglement entropy $S_A$ associated with a subregion $A$ is calculated by the two-point function of twist operators inserted at the boundary of $A$. When the subregion is nearly maximal, that is when the insertions of twist operators are close to the interface, the entanglement entropy $S_A$ does not probe the information of the interface. Then $S_A$ is supposed not to probe it when the size of subregion further decreases, since the insertions of twist operators are far from the interface whose contribution to the two point function is further suppressed due to the locality of interface.

One may try to fix this point by removing the `smoothness' condition. However, as we will see in other examples, one still may not obtain the universal term $\pi c/6$ as the real part in the entanglement entropy so the relaxing does not save it neither.

Another issue with \textbf{Prescription 1} in our dS\textsubscript{3}/ICFT\textsubscript{2} is that it disagrees with the results in AdS\textsubscript{3}/ICFT\textsubscript{2} \cite{Anous:2022,Wei:2023} after analytic continuation. If we take a subregion of size $2\theta$ lying entirely on the asymptotic boundary of the AdS spacetime with smaller radius, the entanglement entropy reads
\begin{equation}\label{eq:AdS/ICFT_EE}
    S_A=\frac{c_+^{\text{AdS}}}{3}\log\left(\frac{2}{\epsilon_{\text{AdS}}}\sin\theta\right) \,.
\end{equation}
where $c_+^{\text{AdS}}$ denotes the central charge of the CFT dual to the AdS spacetime with a smaller radius.
After the analytic continuation $c_+^{\text{AdS}}\to i\tilde{c}_+$ and $\epsilon_{\text{AdS}}\to i\epsilon$, we see that equation (\ref{eq:AdS/ICFT_EE}) becomes
\begin{equation}\label{eq:analytic_continue_SA}
    S_A=i\frac{\tilde{c}_+}{3}\log\left(\frac{2}{\epsilon}\sin\theta\right)+\frac{\pi}{6}\tilde{c}_+
\end{equation}
whose real part does not probe any information of the interface and disagrees with the results obtained by \textbf{Prescription 1} in the dS/ICFT case. 

We do not consider this mismatch as an inconsistency, since the bulk spacetime in the dS/ICFT case cannot be obtained by analytic continuation from that of AdS/ICFT. Therefore, unlike in the pure dS case, we should not expect the results would to agree after analytic continuation. However, we will see that \textbf{Prescription 2} gives a result consistent with the locality of the interface and exactly matches equaion (\ref{eq:analytic_continue_SA}).

We now turn to study \textbf{Prescription 2} \cite{FujikiKoharaShinmyoSuzuki:2025}. The integral that computes the geodesic length is explicitly given by 
\begin{align}\label{eq:geodesic_length}
    L&=l\int\frac{d\rho}{\cos T}\sqrt{-\left(\frac{dT}{d\rho}\right)^2+1} 
    =il\int d\rho\,\frac{\sin\frac{\rho_2-\rho_1}{2}\cos\frac{\rho_2-\rho_1}{2}}{\cos^2\frac{\rho_2-\rho_1}{2}-\sin^2\left(\rho+\frac{\pi}{2}-\frac{\rho_1+\rho_2}{2}\right)}\,.
\end{align}
\textbf{Prescription 2} \cite{FujikiKoharaShinmyoSuzuki:2025} suggests that after integrating along one of the disconnected timelike geodesics corresponding to an integration contour along the real axis, one should extend the integral contour into the imaginary direction and connect it to the other timelike geodesic at $i\infty$, as shown in Fig.\ref{Fig:integral_contour1}. In nother words,
\begin{equation}\label{eq:complex_contour}
    L=i l_+ \left(\int_{\rho_1-\delta}^{\frac{\rho_1+\rho_2}{2}-\frac{\pi}{2}} + \int_{\frac{\rho_1+\rho_2}{2}-\frac{\pi}{2}}^{\frac{\rho_1+\rho_2}{2}-\frac{\pi}{2}+i\infty} + \int_{\frac{\rho_1+\rho_2}{2}+\frac{\pi}{2}+i\infty}^{\frac{\rho_1+\rho_2}{2}+\frac{\pi}{2}} + \int_{\frac{\rho_1+\rho_2}{2}+\frac{\pi}{2}}^{\rho_2+\delta}\right) 
    d\rho\,(\ldots)
\end{equation}
where the dots represent the same integrand as in equation (\ref{eq:geodesic_length}) and $\delta$ denotes the UV cutoff expressed in terms of the $\rho$ coordinate. In equations (\ref{eq:geodesic_length}) and (\ref{eq:complex_contour}), $\rho_1$ and $\rho_2$ denote the endpoints of the interval, and $\frac{\rho_1+\rho_2}{2}$ and $\frac{\rho_2-\rho_1}{2}$ denote the center of the subregion and half the size of the subregion, respectively.
In our case, recall that the subregion is chosen as in equation (\ref{eq:subregion1}), then
\begin{equation}
    \frac{\rho_1+\rho_2}{2}=\pi \,, \quad
    \frac{\rho_2-\rho_1}{2}=\theta\,.
\end{equation}
Before performing the integral, we need to examine the integral contour in the complex $\rho$-plane more carefully. There is a concern about the possible intersection between the geodesic and the brane after complexification of the coordinates\footnote{The author thanks T.Takayanagi for pointing out this issue.}. To analyze this problem, we eliminate the time $T$ from the equations describing the geodesic and the brane and obtain an equation that only depends on $\rho$. We generalize the resulting equation by extending $\rho$ to the full complex plane. A real or complex solution to that equation indicates a possible intersection between the geodesic and the brane, but it is also possibly a mathematical redundancy, depending on the geometric meaning of such solution. 
In our case, eliminating the time $T$ gives  
\begin{equation}\label{eq:looking_for_intersection1}
    \cos^2\rho=\left(\frac{1}{1-R_0^2/l_+^2}+\frac{1}{\cos^2\theta}\right)^{-1}\,.
\end{equation}
The right-hand side of this equation is always positive and smaller than one, so there are no complex solutions for $\rho$. The equation has a real solution, but it is just a mathematical redundancy due to the invariance of $\cos^2(\rho)$ under the reflection $\rho\to\pi-\rho$. As shown in Fig.\ref{Fig:A_fake_vac}, the timelike geodesic does not intersect the brane and the real solution of equation (\ref{eq:looking_for_intersection1}) represents the intersection between the geodesic and the brane only after applying the reflection $\rho\to\pi-\rho$. Therefore, we do not need to worry about any implicit intersection between the geodesic and the brane even after extending the integral contour to the complex plane.

Performing the integral gives us the geodesic length 
\begin{equation}
    L=2il_+\log\left(\frac{2}{\epsilon}\sin\theta\right)
    +\pi l_+
\end{equation}
and the entanglement entropy
\begin{equation}
    S_A=\frac{L}{4G}=i\frac{\tilde{c}_+}{3}\log\left(\frac{2}{\epsilon}\sin\theta\right)+\frac{\pi}{6}\tilde{c}_+\,.
\end{equation}
The imaginary part comes from the contour lying on the real axis while the real part comes from the pole crossing at the endpoints of the interval.
We see that \textbf{Prescription 2} exactly agrees with the results in AdS/ICFT \cite{Anous:2022,Wei:2023} after analytic continuation.

\subsection{Subregion centered at true vaccum}
Due to the inconsistencies of \textbf{Prescription 1} discussed in the previous subsection, for other choices of the subregion $A$ we will focus on \textbf{Prescription 2} and only comment on the issues that would arise if \textbf{Prescription 1} were applied.

Consider a subregion whose center coincides with that of the true vacuum,
\begin{equation}\label{eq:subregion2}
    A=\{0\leq\rho_-\leq\theta,\,\phi=0\}\cup
    \{0\leq\rho_-\leq\theta,\,\phi=\pi\}
\end{equation}
where $\theta\in[0,\frac{\pi}{2}]$.
In this case, the two disconnected timelike geodesics also tend to end at $\rho=\frac{\pi}{2}$ on the $T=0$ slice which lies in the region of the fake vacuum, so the timelike geodesic inevitably intersect the brane. Therefore, the geodesic lying in the region of the true vaccum constitutes only part of the full geodesic. 
To find the entire geodesic, we need to use the junction condition at the brane to find the corresponding intersection point in the region of the fake vacuum, and determine the expression for the segment extending in the fake vacuum by matching the momentum component tangent to the brane. 

\begin{figure}[h]
  \begin{minipage}[b]{0.5\columnwidth}
    \centering
    \includegraphics[width=3.7cm]{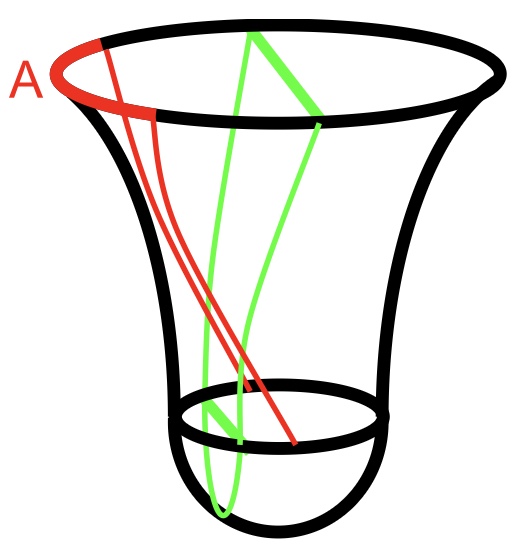}
    \caption{Subregion centered at the center of the true vacuum. The timelike geodesic is shown in red and intersects with the brane which is shown in green.}
    \label{Fig:A_true_vac}
  \end{minipage}
  \hspace{0.04\columnwidth} 
  \begin{minipage}[b]{0.47\columnwidth}
    \centering
    \includegraphics[width=5cm]{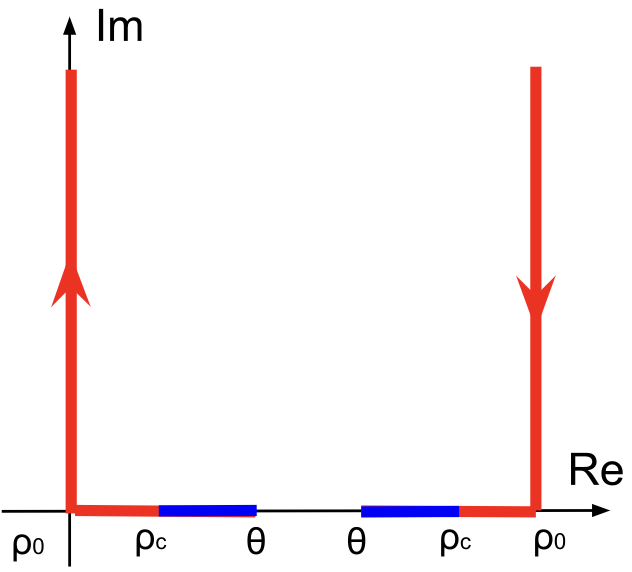}
    \caption{The integral contour of $\rho$. The blue line represents the part in the true vacuum and should be considered as lying in the complex $\rho_-$-plane. The red line represents the part in the fake vacuum and lies in the complex $\rho_+$-plane.}
    \label{Fig:integral_contour2}
  \end{minipage}
\end{figure}

Denote the intersection point viewed in the fake vacuum by $(T_+,\rho_+)=(T_c,\rho_c)$ and the endpoint at the global time $T=0$ by $(T_+,\rho_+)=(0,\rho_0)$. Then the geodesic connecting the two points is given by
\begin{equation}
    \sin(\rho_+-\rho_0)
    =\frac{\sin(\rho_c-\rho_0)}{\sin T_c}\sin T_+\,.
\end{equation}
The conditions which connect this to the segment lying in the true vaccum are then 
\begin{align}
    &l_+\,\tan T_c=\frac{\sqrt{l_-^2-R_0^2}}{\cos\theta} 
    \,;\label{eq:1st_conjunction}\\[3pt] 
    &\frac{l_+\,\sin T_c\,\sin\rho_0}{\sqrt{\cos^2(\rho_c-\rho_0)-\cos^2T_c}}=\frac{l_-}{\sin\theta}
    \,;\label{eq:matching_momentum}\\[3pt] 
    &\cos\rho_c=\sqrt{1-\frac{R_0^2}{l_+^2}}\cos T_c
    \,. \label{eq:geodesic_intersect_brane}
\end{align}
Equation (\ref{eq:1st_conjunction}) comes from the 1\textsuperscript{st} Israel junction condition which says that the induced metric on the brane should be continuous. Equation (\ref{eq:matching_momentum}) matches the momentum component tangent to the brane of the two geodesics. Equation (\ref{eq:geodesic_intersect_brane}) is obtained by noticing that $(T_c,\rho_c)$ is also a point on the brane.
The expressions for $T_c$ and $\rho_c$ is immediately obtained as
\begin{equation}
    \sin T_c=\sqrt{\frac{l_-^2-R_0^2}{l_+^2\cos^2\theta+l_-^2-R_0^2}} 
    \,, \quad
    \sin\rho_c=\sqrt{\frac{l_-^2-R_0^2\sin^2\theta}{l_+^2\cos^2\theta+l_-^2-R_0^2}}
\end{equation}
or equivalently 
\begin{equation}
    \cos T_c=\frac{l_+\,\cos\theta}{\sqrt{l_+^2\cos^2\theta+l_-^2-R_0^2}}
    \,, \quad
    \cos\rho_c=\sqrt{\frac{l_+^2-R_0^2}{l_+^2\cos^2\theta+l_-^2-R_0^2}}\,\cos\theta \,.
\end{equation}
The endpoint $\rho_0$ is determined by 
\begin{equation}\label{eq:rho_0}
    \tan\rho_0=\frac{l_-\cos\theta\left(l_-\sqrt{l_+^2-R_0^2}+l_+\sqrt{l_-^2-R_0^2}\right)}{\sqrt{l_-^2-R_0^2\sin^2\theta}\,(l_+^2-l_-^2)}
\end{equation}
which is negative definite because $l_+<l_-$. Since $\rho_0\in[0,\pi]$, it follows that $\rho_0>\frac{\pi}{2}$, which implies that the endpoints of the timelike geodesics are shifted away from $\frac{\pi}{2}$ due to their intersection with the brane, see Fig.\ref{Fig:A_true_vac}.

The goedesic length includes two parts. One comes from the timelike geodesic lying in the true vacuum, and the other comes from the timelike geodesic lying in the fake vacuum and its extension to the compelx plane, see Fig.\ref{Fig:integral_contour2}.
The part lying in the true vacuum, that is the part shown in blue in Fig.\ref{Fig:integral_contour2}, has length
\begin{align}
    L_{\text{ture vac}}
    &=2il_-\int_{\theta+\delta}^{\tilde{\rho}} d\rho \,
    \frac{\sin\theta\,\cos\theta}{\cos^2\theta-\cos^2\rho}
    \notag \\
    &=2il_-\left[-\tanh^{-1}\left(\frac{\tan\rho}{\tan\theta}\right)\right]_{\theta+\delta}^{\tilde{\rho}}
    \notag \\
    &=2il_-\log\left(\frac{2}{\epsilon}\sin\theta\right)
    -il_-\log\frac{\sqrt{l_-^2-R_0^2\sin^2\theta}+\sin\theta\sqrt{l_-^2-R_0^2}}{\sqrt{l_-^2-R_0^2\sin^2\theta}-\sin\theta\sqrt{l_-^2-R_0^2}} + \pi l_-
\end{align}
where $\tilde{\rho}$ denotes the intersection point between the geodesic
\begin{equation}
    \cos\rho_-=\cos\theta \sin T_-
\end{equation}
and the trajectory of the brane
\begin{equation}
    \cos\rho_-=\sqrt{1-\frac{R_0^2}{l_-^2}}\cos T_-\,.
\end{equation}

Again, we need to examine whether the extension of the integration contour to the complex plane leads to any implicit intersection with the brane. Eliminating the time coordinate $T_+$ from the equations describing the geodesic and the brane gives
\begin{equation}
    \frac{\cos^2\rho}{1-\frac{R_0^2}{l_+^2}}+\frac{\sin^2T_c}{\sin^2(\rho_c-\rho_0)}\sin^2(\rho-\rho_0)=1 \,.
\end{equation}
Since the resulting equation already has a real solution $\rho=\rho_c$ and all the coefficients in the equation above are real, it no longer admits any complex solutions for $\rho$. Therefore, the integral over the segment in the fake vaccum can be extended to the complex plane without concern about additional intersections with the brane.

The segment in the fake vacuum has length
\begin{align}
    L_{\text{fake vac}}
    &=il_+\left(\int_{\rho_c}^{\rho_0}+\int_{\rho_0}^{\rho_0+i\infty}+\int_{\rho_0+\pi+i\infty}^{\rho_0+\pi}+\int_{\rho_0+\pi}^{\rho_c+\pi}\right)d\rho \,
    \frac{\sin(\rho_0-\rho_c)\sqrt{\sin^2T_c-\sin^2(\rho_0-\rho_c)}}{\sin^2(\rho_0-\rho_c)-\sin^2(\rho-\rho_)\sin^2T_c} 
    \notag \\
    &=2il_+\,\tanh^{-1}\left(\frac{\sqrt{\cos^2(\rho_0-\rho_c)}-\cos^2T_c}{\cos(\rho_0-\rho_c)}\right)  \notag\\
    &=2il_+\tanh^{-1}\left(\frac{1}{X}\right) 
\end{align}
where we have substituted the expression for $\rho_0,\rho_c$ and $T_c$ in the last equality and defined
\begin{equation}
    X\coloneq\frac{\cos(\rho_c-\rho_0)}{\sqrt{\cos^2(\rho_c-\rho_0)-\cos^2T_c}}
    =\frac{\sqrt{l_-^2-R_0^2\sin^2\theta}}{R_0^2\sin\theta}
    \left(\frac{l_+\,l_-}{\sqrt{l_-^2-R_0^2}}-\sqrt{l_+^2-R_0^2}\right)\,.
\end{equation}
The entropy associated with the subregion (\ref{eq:subregion2}) is then 
\begin{align}\label{eq:S_A2}
    S_A=&\frac{1}{4G}(L_{\text{true vac}}+L_{\text{fake vac}})
    \notag \\
    =&i\frac{\tilde{c}_-}{3}\log\left(\frac{2}{\epsilon}\sin\theta\right) +\frac{\pi}{6}\tilde{c}_- \notag \\
    &-i\frac{\tilde{c}_-}{6}\log\frac{\sqrt{l_-^2-R_0^2\sin^2\theta}+\sin\theta\sqrt{l_-^2-R_0^2}}{\sqrt{l_-^2-R_0^2\sin^2\theta}-\sin\theta\sqrt{l_-^2-R_0^2}} 
    +i\frac{\tilde{c}_+}{6}\log\frac{X+1}{X-1}
\end{align}

where we have used $\tanh^{-1}x=\frac{1}{2}\log\frac{1+x}{1-x}$.
From this expression, we see that the UV-divergent term and the real part remain unchanged in the presence of the brane. The modification due to the presence of the brane appears only in  the $O(1)$ imaginary term which comes from the truncation of the segment in the true vacuum and the shift in the endpoints of the segment in the fake vacuum.

We note that our dS/ICFT result (\ref{eq:S_A2}) shares some similarities with that of the AdS/ICFT case \cite{Anous:2022}. In both cases, the UV-divergent terms is detemined by the side on which the subregion is chosen. Indeed, the first line of (\ref{eq:S_A2}) matches the UV-divergent term in the AdS/ICFT case \cite{Anous:2022} after analytic continuation. In both cases, the deformation of the entanglement entropy $S_A$ is encoded in a logarithmic function, though their explicit expressions are not related by analytic continuation. 

A notable distinction is that, in our case, the geodesic always intersects the brane, as long as the subregion $A$ is chosen to be completely lying in the true vacuum. However, in the AdS/ICFT case, one can choose the subregion to be far from the interface or choose it to be sufficiently small so that the geodesic does not intersect the brane \cite{Anous:2022}. Such a distinction reveals an unexpected feature of the CFT dual to the de Sitter spacetime and deserves further study.

We also comment on what would happen if \textbf{Prescription 1} were applied. The timelike part remains the same and the only difference comes from the spacelike part. For two points on the rim of a hemisphere, a geodesic connecting them can extend into the interior of the hemisphere only if the two points are antipodal \cite{Doi:2023JHEP}. In our case, as discussed below equation (\ref{eq:rho_0}), the endpoints are located at $\rho=\rho_0$ which larger than $\frac{\pi}{2}$ and thus are not antipodal. Therefore, the only space-like geodesic that can connect the two points is an arc along the rim, whose length does not reproduce the universal $\frac{\pi}{6}\tilde{c}$ term.

\subsection{Subregion perpendicular to interface and $g$-function}
As a final test, let us consider a subregion centered at the interface 
\begin{equation}\label{eq:subregion3}
    A=\{\tfrac{\pi}{2}-\theta\leq \rho_- \leq \tfrac{\pi}{2}\,,\phi=0\} \cup
    \{\tfrac{\pi}{2}\leq \rho_+ \leq\tfrac{\pi}{2}+\theta\,,\phi=0\}
\end{equation}
where $\theta\in[0,\frac{\pi}{2}]$. 
Since the geodesic consists of two segments separated by the brane, the integration constants associated with the two segments should be treated independently and should be determined by junction conditions at the brane, as in the previous subsection. We write the geodesics as 
\begin{equation}
    \sin(\rho_\pm-\rho_0^{(\pm)})=\sin(\frac{\pi}{2}\pm\theta-\rho_0^{(\pm)})\,\sin T_\pm 
\end{equation}
where $\pm$ labels the geodesics on the two sides of the brane and $\rho_0^{(\pm)}$ denotes the endpoints of the two geodesics on the $T=0$ slice. The equations that determine $\rho_0^{(\pm)}$ and the intersection points are 
\begin{align}
    &\frac{\cos^2\rho_-}{1-R_0^2/l_-^2}+\frac{\sin^2(\rho_--\rho_0^{(-)})}{\sin^2(\tfrac{\pi}{2}-\theta-\rho_0^{(-)})}=1 
    \label{eq:perpendicular_1}\\
    &\frac{\cos^2\rho_+}{1-R_0^2/l_+^2}+\frac{\sin^2(\rho_+-\rho_0^{(+)})}{\sin^2(\tfrac{\pi}{2}+\theta-\rho_0^{(+)})}=1 
    \label{eq:perpendicular_2}\\
    &\frac{\sqrt{l_-^2-R_0^2}}{\sin(\tfrac{\pi}{2}-\theta-\rho_0^{(-)})}\frac{\sin(\rho_--\rho_0^{(-)})}{\cos\rho_-}=
    \frac{\sqrt{l_+^2-R_0^2}}{\sin(\tfrac{\pi}{2}+\theta-\rho_0^{(+)})}\frac{\sin(\rho_+-\rho_0^{(+)})}{\cos\rho_+} 
    \label{eq:perpendicular_3}\\
    &\frac{l_- \sin\rho_0^{(-)}}{\cos(\tfrac{\pi}{2}-\theta-\rho_0^{(-)})}=
    \frac{l_+\sin\rho_0^{(+)}}{\cos(\tfrac{\pi}{2}+\theta-\rho_0^{(+)})}
    \label{eq:perpendicular_4}
\end{align}
where the first two come from eliminating the time $T_\pm$ from the equations describing the geodesics and the branes, the third equation follows from the continuity of the induced metric on the brane and the last one matches the momentum component tangent to the brane. Note that we have eliminated the time coordinates $T_\pm$ and written all the equations in terms of $\rho_\pm$ so that they can be easily extended to the complex plane. Equations (\ref{eq:perpendicular_1}-\ref{eq:perpendicular_4}) admit real solutions, but such solutions indicate that the timelike geodesic closes on the brane and is completely timelike, and thus the absence of a $g$-function. To understand why this is regarded as a missing of $g$-function, we first recall that, in the AdS/ICFT case \cite{Takayanagi:2007,Anous:2022,Wei:2023}, if the subregion is chosen as in (\ref{eq:subregion3}), the entanglement entropy is given by
\begin{equation}
    S_A=\frac{c_1+c_2}{6}\log\left(\frac{2}{\epsilon}\sin\theta\right)
    +\frac{c_1}{6}\xi_*^{(1)}+\frac{c_2}{6}\xi_*^{(2)}
\end{equation}
where $c_1$ and $c_2$ denote the central charges of the two CFTs joined by the interface and $\xi_*^{(1)},\xi_*^{(2)}$ are constants determined by the tension and the AdS radii. The last two terms originate from the intersection of the geodesic with the brane, and are therefore recognized as $g$-function \cite{Affleck:1991,CalabreseCardy:2004}. Since the expressions $l_1\cdot\xi_*^{(1)}$ and $l_2\cdot\xi_*^{(2)}$ remain real under the analytic continuation $l_{1,2}\to il_{1,2}$, one would expect a $g$-function term to appear in the real part of $S_A$ in the dS/ICFT case. However, for the completely timelike geodesic, the real part of $S_A$ arises from the analytic continuation of the UV cutoff $\epsilon$ by $\epsilon\to i\epsilon$, and therefore has nothing to do with the $g$-function.

\begin{figure}[h]
  \begin{minipage}[b]{0.5\columnwidth}
    \centering
    \includegraphics[width=4cm]{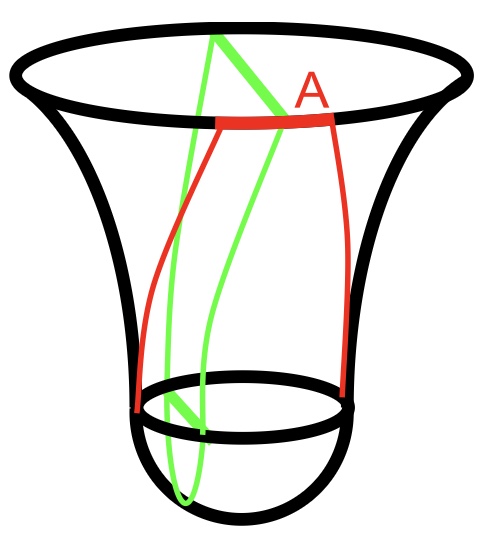}
    \caption{Subregion centered centered at the interface. The time-like geodesic is shown in red and never intersects with the brane which is shown in green.}
    \label{Fig:A_cross_brane}
  \end{minipage}
  \hspace{0.04\columnwidth} 
  \begin{minipage}[b]{0.47\columnwidth}
    \centering
    \includegraphics[width=5cm]{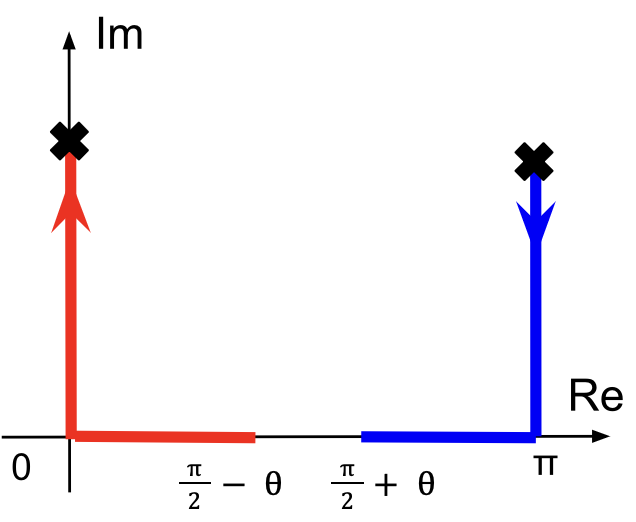}
    \caption{The integral contour of $\rho_\pm$. The blue line represents the part in the fake vacuum and the red line represents the part in the true vacuum, and they should be considered as lying on different complex planes. The black cross denotes the intersection points.}
    \label{Fig:integral_contour3}
  \end{minipage}
\end{figure}

To obtain the $g$-function, we need to consider complex solutions of equations (\ref{eq:perpendicular_1}-\ref{eq:perpendicular_4}). The only way to do this is to make equation (\ref{eq:perpendicular_4}) identically satisfied, namely by setting $\rho_0^{(-)}=0$ and $\rho_0^{(+)}=\pi$, as shown in Fig.\ref{Fig:A_cross_brane}. The two segments of the timelike geodesic are then described by
\begin{align}
    &\sin\rho_-=\cos\theta\,\sin T_- \,; 
    \label{eq:true_vac_geodesic}\\
    &\sin\rho_+=\cos\theta\,\sin T_+ \,. 
    \label{eq:fake_vac_geodesic}
\end{align}
where $T_-,T_+\in[0,\frac{\pi}{2}]$, $\rho_-\in[0,\frac{\pi}{2}]$ and $\rho_+\in[\frac{\pi}{2},\pi]$.
In this case, equation (\ref{eq:perpendicular_3}) reduces identically $iR_0/\sin\theta$ after substituting equations (\ref{eq:perpendicular_1}) and (\ref{eq:perpendicular_2}). Moreover, equations (\ref{eq:perpendicular_1}) and (\ref{eq:perpendicular_2}) can be simplified to
\begin{equation}
    \cos^2\rho_\pm= \frac{\cos^2\theta-1}{\frac{\cos^2\theta}{1-R_0^2/l_\pm^2}-1}\,.
\end{equation}
The right-hand side is either negative or greater than one, so the solution for $\rho$ is always complex. The solution, though depending on the value of $\theta$, is given by
\begin{equation}\label{eq:complex_intersection}
    \rho_\pm = \left\{ 
        \begin{aligned}
            &\frac{\pi}{2}+i\sinh^{-1}\sqrt{\frac{1-\cos^2\theta}{\frac{\cos^2\theta}{1-R_0^2/l_\pm^2}-1}}+n\pi \,,
            &\quad&0\leq \theta\leq\cos^{-1}\sqrt{1-\frac{R_0^2}{l_\pm^2}} \\
            &i\cosh^{-1}\sqrt{\frac{\cos^2\theta-1}{\frac{\cos^2\theta}{1-R_0^2/l_\pm^2}-1}}+ n\pi \,,
            &\quad&\cos^{-1}\sqrt{1-\frac{R_0^2}{l_\pm^2}}\leq\theta\leq\frac{\pi}{2}
        \end{aligned}
     \right.
\end{equation}
where $n\in\mathbb{Z}$. The trajectory of this intersection point in the complex plane as $\theta$ increases is described as follows. This point starts from $\frac{\pi}{2}+n\pi$ when $\theta=0$ and moves in the imaginary direction while keeping the real part fixed. At a critical value of $\theta$, it reaches $i\infty$ and its real part changes to $n\pi$. Then it moves backward until its imaginary part reaches a finite value $\cosh^{-1}1$.
\\

To compute the geodesic length, one should divide the integral into an integral over (\ref{eq:true_vac_geodesic}) and that over (\ref{eq:fake_vac_geodesic}) and extend the corresponding integration contours to the intersection points in the complex plane, see Fig.\ref{Fig:integral_contour3}. Surprisingly, though there is a large redundancy which is by no means resolved and a $\theta$-dependence in the solution (\ref{eq:complex_intersection}), the contribution from the intersection point does not depend on the choice of $n$ and $\theta$.
\begin{align}
    L=&il_-\left(\int_{\frac{\pi}{2}-\theta}^0+\int_0^{\rho_-}\right)d\rho \, \frac{\sin\theta\,\cos\theta}{\cos^2\theta-\sin^2\rho} 
    +il_+\left(\int_{\rho_+}^\pi+\int_{\pi}^{\frac{\pi}{2}+\theta}\right)d\rho \, \frac{\sin\theta\,\cos\theta}{\cos^2\theta-\sin^2\rho}  \notag \\
    =&i(l_+ + l_-)\log\left(\frac{2}{\epsilon}\sin\theta\right)
    +\frac{\pi}{2}(l_++l_-)
    +l_+\tan^{-1}\frac{-R_0}{\sqrt{l_+^2-R_0^2}}
    +l_-\tan^{-1}\frac{R_0}{\sqrt{l_-^2-R_0^2}}\,.
\end{align}
where the arctangent function takes values from $0$ to $\pi$. Then the entropy $S_A$ is computed as 
\begin{align}\label{eq:SA_3}
    S_A=\frac{L}{4G}
    =&i\frac{\tilde{c}_++\tilde{c}_-}{6}
    \log\left(\frac{2}{\epsilon}\sin\theta\right)
    +\frac{\pi}{12}(\tilde{c}_++\tilde{c}_-) \notag \\
    &+\frac{\tilde{c}_+}{6}\tan^{-1}\frac{-R_0}{\sqrt{l_+^2-R_0^2}}
    +\frac{\tilde{c}_-}{6}\tan^{-1}\frac{R_0}{\sqrt{l_-^2-R_0^2}}
\end{align}
The $g$-function is read off as 
\begin{equation}\label{eq:g-function}
    \log g
    =\frac{\tilde{c}_+}{6}\tan^{-1}\frac{-R_0}{\sqrt{l_+^2-R_0^2}}
    +\frac{\tilde{c}_-}{6}\tan^{-1}\frac{R_0}{\sqrt{l_-^2-R_0^2}}\,.
\end{equation}

As a final remark, we comment on \textbf{Prescription 1}. If we were to apply \textbf{Prescription 1} in this case, we would not get the $\frac{\pi}{12}(\tilde{c}_++\tilde{c}_-)$ term in equation (\ref{eq:SA_3}), regardless whether the smoothness condition is imposed. One may also try to interpret the deviation from $\frac{\pi}{12}(\tilde{c}_++\tilde{c}_-)$ in the real part. However, with the smoothness condition imposed, the intersection point of the spacelike geodesic and the brane moves as the size of the subregion varies. The $\theta$-dependence makes it inappropriate for such a term to be considered as a $g$-function. If the smoothness condition is removed, the spacelike geodesic intersects the brane orthogonally and indeed reproduces the same $g$-function (\ref{eq:g-function}), but the $\frac{\pi}{12}(\tilde{c}_++\tilde{c}_-)$ term is then missing.

\section{Conclusion and outlook}\label{conclusion}
In this paper, we have established the dS\textsubscript{d+1}/ICFT\textsubscript{d} correspondence through a bottom-up approach in which the Hartle-Hawking state \cite{HartleHawking:1983} is replaced by what we call Coleman-De Luccia state \cite{Coleman-DeLuccia:1980} in the dictionary. The CDL state $\Psi_{\text{CDL}}$ is obtained as a Euclidean solution of the Einstein equation in the scenario of vacuum decay under the thin-wall approximation. After absorbing the inflaton potential by a brane with tension $\sigma$, the CDL state can be treated as two portions of spheres with different radii glued at the brane. We found that, due to the unconventional coordinates used in the Euclidean solution, the desired Lorentzian spacetime cannot be obtained by analytic continuation. The Lorentzian spacetime tailored for the dS/ICFT correspondence is obtained by working out the Hamiltonian formalism \cite{Cespedes:2020}.
The bulk spacetime thus consists of two de Sitter spacetimes with different radii joined by a brane that traces out a dS slice. Since each of the de Sitter spacetimes is conjectured to be dual to a non-unitary Euclidean CFT \cite{Strominger:2001,Maldacena:2002}, the bulk spcacetime is supposed to be dual to two non-unitary CFTs joined by an interface.

We used the established dS/ICFT correspondence to test two prescriptions for extremal surfaces. One proposes that the timelike geodesic in the Lorentzian signature should be smoothly extended to a spacelike geodesic in the Euclidean signature \cite{Narayan:2022}, and the other suggests that it should be extended to the imaginary direction of compelx plane \cite{FujikiKoharaShinmyoSuzuki:2025}. 
We found that \textbf{Prescription 1} will lead to a contradiction with the results expected from the CFT side. Even if the smoothness condition is removed, the spacelike geodesic still does not produce physically sensible results for generic subregions.
Whereas, \textbf{Prescription 2} works well and reveals similarities and differences between our dS/ICFT and AdS/ICFT \cite{Takayanagi:2007,Anous:2022,Wei:2023}. We also obtained the $g$-function encoding information of the interface \cite{Affleck:1991,CalabreseCardy:2004} by considering a subregion perpendicular to the interface, as expected. The properties of this $g$-function deserves further study. 

As a final remark, we note a recent work \cite{Guo2026} that attempts to select the optimal prescription for extremal surfaces in pure dS by imposing the Kontsevich-Segal-Witten criterion \cite{KontsevichSegal:2021,Witten:2021} in the holomorphic complex geometry\cite{LeBrun1983SpacesOC}. This criterion is quite reasonable from the perspective of the gravitational path integral over replicated geometries. However, the resulting prescription suggests that the extremal surface in pure dS consists of a timelike geodesic truncated at the cosmological horizon and a spacelike geodesic in another Lorentzian spcacetime glued across the cosmological horizon\footnote{The author thanks Wu-zhong Guo for a discussion on his work.}. 
Therefore, this prescription has a geometric interpretation entirely different from those of \textbf{Prescription 1} and \textbf{Prescription 2} studied in this paper.

We hope that further studies of the extremal surfaces either in the complexified coordinates or in the holomorphic complex geometry can tell us more about the entropy in de Sitter spacetime.

\section*{Acknowledgment}
The author thanks K.Narayan, T.Takayanagi and W.Guo for useful discussions.
The author especially thanks T.Takayanagi for a first reading of this manuscript and suggestions.
The author thanks RIKEN iTHEMS for several discussions and comments received during the workshop ``de Sitter Holography Meets Non-Hermitian Quantum Matter".

\bibliographystyle{JHEP}
\bibliography{references}

\end{document}